# The Intracluster Light and its Link with the Dynamical State of the Host Group/Cluster: the Role of the Halo Concentration

Emanuele Contini,[1] Seyoung Jeon,[1] Jinsu Rhee,[1,2] San Han,[1] and Sukyoung K. Yi[1]

[1]*Department of Astronomy and Yonsei University Observatory, Yonsei University, 50 Yonsei-ro, Seodaemun-gu, Seoul 03722, Republic of Korea*

[2]*Korea Astronomy and Space Science Institute, 776, Daedeokdae-ro, Yuseong-gu, Daejeon 34055, Republic of Korea*

## ABSTRACT

We investigate on the role of the halo concentration in the formation of the intra-cluster light (ICL) in galaxy groups and clusters, as predicted by a state-of-art semi-analytic model of galaxy formation, coupled with a set of high-resolution dark matter only simulations. The analysis focuses on how the fraction of ICL correlates with halo mass, concentration and fraction of early-type galaxies (ETGs) in a large sample of groups and clusters with $13.0 \leq \log M_{halo} \leq 15.0$. The fraction of ICL follows a normal distribution, a consequence of the stochastic nature of the physical processes responsible for the formation of the diffuse light. The fractional budget of ICL depends on both halo mass (very weakly) until group scales, and concentration (remarkably). More interestingly, the ICL fraction is higher in more concentrated objects, a result of the stronger tidal forces acting in the innermost regions of the haloes where the concentration is the quantity playing the most relevant role. Our model predictions do not show any dependence between the ICL and ETGs fractions and so, we instead suggest the concentration rather than the mass, as recently claimed, to be the main driver of the ICL formation. The diffuse light starts to form in groups via stellar stripping and mergers and later assembled in more massive objects. However, the formation and assembly keep going on group/cluster scales at lower redshift through the same processes, mainly via stellar stripping in the vicinity of the central regions where tidal forces are stronger.

## 1. INTRODUCTION

Galaxy groups and clusters host galaxies with different morphologies that mainly depend on their histories and for how long they have been experiencing the two most important quenching mechanisms, i.e. mass and environmental quenching (Peng et al. 2012; Contini et al. 2017a; Kawinwanichakij et al. 2017; Davies et al. 2019; Rhee et al. 2020; Contini et al. 2020; Mao et al. 2022). These objects, not only host galaxies with different properties, but a large part of the stellar content within them is made by a diffuse light from stars that are not bound to any galaxy, and feel only the potential well of the dark matter halo. This diffuse light is usually called intracluster light (ICL, Contini 2021; Montes 2022; Arnaboldi &

Gerhard 2022). Since its first discovery (Zwicky 1937), the ICL has been studied as an important stellar component of galaxy groups and clusters, but only in the most recent years, given the modern technologies, the scientific community has made remarkable steps forward in understanding its formation and evolution, on both the observational and theoretical sides (Murante et al. 2007; Rudick et al. 2009; Contini et al. 2014; Burke et al. 2015; Mihos et al. 2017; Morishita et al. 2017; DeMaio et al. 2018; Montes & Trujillo 2018; Zhang et al. 2019; Iodice et al. 2021; Ragusa et al. 2021; Joo & Jee 2023, and many others).

One weak point in studying the ICL is its own definition. Being a faint component, it is usually not trivial to detect and, on top of that, it is even more difficult to isolate it from the rest, satellite galaxies in the cluster and the brightest group/cluster galaxy (BGG/BCG). Indeed, observers are able to isolate the ICL by removing

Corresponding author: Emanuele Contini

emanuele.contini82@gmail.com

the contribution given by the sky and by masking the satellite galaxies, but the separation from the BCG, with which it is embedded, is still far from being accurately achieved. One particular difficulty in this case is the separation between the stellar halo, which is also known as the envelope bound to the BCG, with the ICL. Stellar haloes and ICL have a common origin and in most of the cases they cannot be separated using only imaging (Montes 2022 for a review), but for nearby objects, such as Virgo A, it has been possible by using the velocity distribution of Planetary Nebulae (Longobardi et al. 2015). From a theoretical point of view, the two components can be separated by using the dynamical information provided by numerical simulations (e.g., Dolag et al. 2010; Cui et al. 2014), or as a natural consequence of the result of differential equations in semi-analytic models (e.g., Guo et al. 2011; Contini et al. 2014). Observers can only rely on the information coming from the light, and a complete separation will never be possible. Nevertheless, there are mainly two techniques that provide acceptable results: a surface brightness (arbitrary) cut which blindly separates BCG and ICL (e.g., Zibetti et al. 2005), and a more accurate profile fitting able to separate the two components and to identify also the region of transition between the two (e.g., Montes et al. 2021).

Taking advantage of these observational techniques and numerical methods, it has been possible to study the properties of the ICL. Colors, metallicity and age have been the focus of several works (Gonzalez et al. 2013; Montes & Trujillo 2014; DeMaio et al. 2015; Edwards et al. 2016; Iodice et al. 2017; Ko & Jee 2018; Spavone et al. 2018; Contini et al. 2019; Gu et al. 2020; Ragusa et al. 2021; Joo & Jee 2023; Werner et al. 2023), and all of them with the goal of sheding some light on the possible channels for the formation of the ICL.

In the last two decades, thanks to the considerable improvement made on numerical simulations, several mechanisms have been proposed as channels for the formation of the ICL. Based on the average opinion among the scientific community and starting from the least important to close with the main channel, although there is not a definite and satisfactory consensus yet, the ICL may form through: *in-situ* star formation (Puchwein et al. 2010), disruption of dwarf satellites (Conroy et al. 2007; Purcell et al. 2007; Giallongo et al. 2014; Annunziatella et al. 2016; Raj et al. 2020), pre-processing/accretion (Mihos et al. 2005; Sommer-Larsen 2006; Contini et al. 2014; Ragusa et al. 2023), minor/major mergers (Monaco et al. 2006; Murante et al. 2007; Burke et al. 2015; Groenewald et al. 2017; Contini et al. 2014; Joo & Jee 2023), and stellar stripping of intermediate/massive satellite galaxies (Rudick et al. 2009, 2011; Contini et al. 2014; DeMaio et al. 2015, 2018; Contini et al. 2018; Montes & Trujillo 2018).

*In-situ* star formation has been ruled out by observational studies which found that the contribution given to the ICL by this channel is around 1% (e.g., Melnick et al. 2012). The disruption of dwarf satellites is important in terms of number of galaxies that can be destroyed during the process of the ICL formation [1], but it is almost negligible in terms of stellar mass deposited in the ICL (e.g., Murante et al. 2007; Contini et al. 2014). Pre-processing/accretion can be a relevant channel, and in particular in massive ($\log M_{halo} > 14.5$) haloes (e.g., Contini et al. 2014; Joo & Jee 2023). This process refers to the amount of stellar mass in ICL that forms in groups and later accreted in another group or cluster through a merger between haloes. As will be discussed below, this process can be considered as a sub-channel because it is a way to increase the ICL content, but not to directly form it. Indeed, the accreted ICL was already formed via other channels.

The suppression of *in-situ* star formation, the negligible contribution given by the disruption of dwarf satellites and the downgrading of pre-processing (important but indirect) leave room for the other two direct mechanisms: violent relaxation during mergers and stellar stripping of intermediate/massive satellites. These two processes together are responsible for the formation of the bulk of the ICL, although it is not straightforward to show which one dominates. In Contini et al. (2014, 2018) we have shown that stellar stripping is the major channel and contributes the most to the ICL, but also argued that without a clear definition of a merger, the two channels can be mostly indistinguishable (see also Contini 2021). A possible way to discern among them is given by the study of colors, age and metallicity of the ICL, without even separate it from the BCG. If mergers are dominant, we would expect no color/metallicity gradient in BCG+ICL systems (e.g., Montes & Trujillo 2018), simply because mergers mix up the stellar populations (e.g., Contini et al. 2019). On the contrary, if more gentle events such as stellar stripping are dominant, we would expect a gradient in colors and metallicity, because the BCG dominated region should be more metal-rich and redder than the region dominated by the ICL (see Contini 2021 for a full discussion on this point), given the fact that stars are coming from the outermost parts of intermediate/massive satellites. In observations, there is evidence for both (e.g., Montes & Trujillo 2018; Joo & Jee 2023), but more numerous examples of gradients at least in the local universe/ low redshift. This essentially implies that stellar stripping, being more gentle than mergers and not mixing up the stellar populations, might be the dominant process, but a single recent merger can cancel out the color/metallicity gradient (Contini et al. 2019) even though stellar stripping contributed the most.

---

[1] This process is important in the context of the stellar mass function. Indeed, it alleviates the conflict between the observed and predicted stellar mass functions on small stellar masses when models consider this channel for the ICL formation (e.g., Annunziatella et al. 2014; Contini et al. 2014, 2017b).

In this work we want to address the following point: considering that stellar stripping and mergers are the responsible processes for the formation of the ICL, our focus is to link the fraction of ICL in a large range of halo masses with the current dynamical state of the halo, which we identify with the halo concentration. Taking profit of a much larger sample of haloes with respect to that used in Contini et al. 2014 (hereafter C14), we will re-address the relation between the ICL fraction and the halo mass. In C14 we have shown that the halo concentration can explain the large scatter seen in that relation for a narrow range of halo mass, in a way that more concentrated haloes have also a larger ICL fraction. Here, we use the same approach, but go deeper in the role played by the concentration.

Recent works have tried to link the evolutionary state of a given object with the fraction of the ICL. Just to mention a few as examples, Iodice et al. (2020) found an increasing fraction of ICL with a decreasing amount of neutral hydrogen, while Ragusa et al. (2023) found a weak increasing fraction of ICL with increasing fraction of early-type galaxies. In particular, these authors found no relation between the ICL fraction and the halo mass, concluding that, when the fraction of early-type galaxies is used as a tracer of the dynamical state of a halo, those more dynamically evolved (larger fraction of early-type galaxies) have also a larger fraction of ICL. They concluded that such correlations can be explained altogether if the bulk of the ICL is formed on group scales and later accreted by clusters. This interpretation is correct and leads to a picture where the mass is driving the formation and assembly of the ICL. However, another possibility is given by a gradual assembly of ICL on both scales, accreted from group scales (which comes from the pre-processing channel) and formed *in-situ* mostly by stellar stripping (or mergers) if the concentration plays an important role. Our goal is to prove that the latter is a more reliable explanation.

The remainder of the paper is structured as follows. In Section 2 we introduce the set of N-body simulations we use as input for our semi-analytic model (SAM), and

**Table 1.** Set of Simulations

| Name | B ($Mpc/h$) | NP | MR ($M_\odot/h$) |
|---|---|---|---|
| YS200 | 200 | $1280^3$ | $3.26 \cdot 10^8$ |
| YS125 | 125 | $1024^3$ | $1.57 \cdot 10^8$ |
| YS75 | 75 | $512^3$ | $2.68 \cdot 10^8$ |
| YS50 | 50 | $512^3$ | $7.96 \cdot 10^7$ |
| YS50HR | 50 | $1024^3$ | $10^7$ |
| YS25HR | 25 | $512^3$ | $10^7$ |

NOTE—First column: name of the simulation; second column: side of the box; third column: total number of DM particles used; fourth column: mass resolution, i.e. the mass of each dark matter particle. It must be noted that the high-resolution runs (HR) have been used only as a test for the convergence of results due to the resolution, and in Figure 6. See text for further details.

further provide a summary of the prescription used in the SAM to implement the formation of the ICL. Section 3 is dedicated to the analysis and to a brief evaluation of the most important results, which will be fully discussed in Section 4 along with the core of the paper, i.e., the link (given by the concentration) between the ICL and the dynamical state of its host. In Section 5 we summarize our main conclusions. In the following, stellar masses are computed with the assumption of a Chabrier (2003) initial mass function and luminosities are computed in AB system. Unless otherwise stated, all units are h-corrected.

## 2. METHODS

In this section we focus on the methodological aspect of our study, by describing the set of numerical simulations we developed (Section 2.1), and used to build the merger trees needed by our SAM as input. Moreover, given the importance in the context of this study, and to avoid referring the reader to several papers, in Section 2.2 we summarize the prescription for the ICL formation implemented in the SAM, by giving particular emphasis to the main features that result necessary for the comprehension of the analysis shown below.

### 2.1. *N-body Simulations*

We take advantage of a set of six dark matter only cosmological simulations carried out by using the latest version of the GADGET-code, i.e. GADGET-4 (Springel et al. 2021). A description of the new improvements brought by the 4th version of GADGET are beyond the scope of this work. Nevertheless, it is worth mentioning that GADGET-4 provides remarkable improve-ments in terms of force accuracy, time-stepping, computational efficiency and more. Not only, a considerable pro with respect to previous versions such as GADGET-3 (Springel 2005), for what concerns SAMs, is the on-the-fly merger tree building offered by GADGET-4.

The main parameters of the set of simulations, such as box size, number of particles and mass resolution are presented in Table 1 [2]. The volume of the boxes goes from $(25\,Mpc/h)^3$ up to $(200\,Mpc/h)^3$. The two smallest boxes have a higher mass resolution (bf more than one order of magnitude with respect to the other boxes), while all the runs share in common the same softening length, $3\,kpc/h$, and the same following Planck 2018 cosmology (Planck Collaboration et al. 2020): $\Omega_m = 0.31$ for the total matter density, $\Omega_\Lambda = 0.69$ for the cosmological constant, $n_s = 0.97$ for the primordial spectral index, $\sigma_8 = 0.81$ for the power spectrum normalization, and $h = 0.68$ for the normalized Hubble parameter. Moreover, all the simulations ran from redshift $z = 63$ to the present time and the data have been stored in 100 discrete snapshots from $z = 20$ to $z = 0$. YS50 and YS75 have also their counterparts where the data have been stored from the same redshifts, but in 50 discrete snapshots in order to investigate on the possible importance of the data storage refinement [3].

Among the 6 different runs, YS25HR and YS50HR are high-resolution simulations. We do not make use of them for most of the analysis (only in Figure 6) given the large sample of haloes that the others already provide. They have been thought and developed to investigate on the convergence of the results in terms of mass resolution. We will come back on this point below in Section 3.

## 2.2. ICL prescription

SAMs run on the merger trees derived from numerical simulations, and their goal is to populate dark matter haloes with galaxies by implementing the physics of baryons. The description of the several physical processes that are implemented in SAMs is beyond the scope of this paper, but it is important to briefly summarize the main features of the processes implemented in our SAM and that bring to the formation of the ICL.

The model follows the original prescription TidRad+Merg described in C14, and further developed in Contini et al. (2018, 2019). This prescription is the combinations of two channels: (1) the tidal interactions between satellite galaxies and the potential well of the host halo (TidRad) and, (2) violent relaxation in galaxy mergers (Merg). The first one is responsible for the stellar stripping, a process that can either disrupt the satellite or strip part of its the stellar mass. In our prescription, the amount of stellar mass stripped is assumed to be accumulated into an extra component, the ICL, thus forming the diffuse light that is observed in groups and clusters of galaxies. The ICL is also considered to be associated with the BCG, in the sense that it surrounds the main galaxy of the host halo, but it is not bound to it. However, there are particular cases for which satellite galaxies can have their own ICL, depending on the potential well of the subhalo associated to them. We will provide more details on this point below.

Mathematically speaking, the stellar density profile of satellite galaxies is approximated by a spherically symmetric isothermal profile, such that the tidal radius $R_t$, i.e. the radius at which the gravity of the satellite is no longer able to keep it stable, can be estimated with the following equation (see Binney & Tremaine 2008):

$$R_t = \left(\frac{M_{sat}}{3 \cdot M_{halo}}\right)^{1/3} \cdot D\,, \qquad (1)$$

where $M_{sat}$ is the satellite mass (stellar mass + cold gas mass), $M_{halo}$ is the dark matter mass of the parent halo, and $D$ the satellite distance from the centre of the halo.

In the computation of the tidal radius the model assumes an isothermal profile, but a more realistic satellite is assumed when tidal stripping is applied to it. Specifically, we consider the satellite galaxy to be a two-component system, where the bulge is modelled with a Jaffe (Jaffe 1983) profile, and the disk is assumed to follow an exponential distribution. When the tidal forces are strong enough so that the tidal radius is smaller than the bulge radius, the satellite is totally disrupted. However, most of the times the satellite's own gravity can be remarkable and the tidal radius reaches the disk of the satellite. In this case (and again it happens in the majority of the cases), only the stellar mass in the shell $R_t - R_{sat}$ is stripped. The amount of stellar mass the satellite has been deprived of, in either case, joins the ICL component associated with the BCG, while a proportional fraction of cold gas is moved to the hot component of the BCG. Right after the stripping event, the tidal radius automatically becomes the radius of the new disk, and given that we assume an exponential pro-

---

file, for which 99.9% of the stellar mass is contained within ten times the scale length of the disk, the new scale length $R_{sl}$ is updated to $R_t/10$.

Satellite galaxies are embedded in subhaloes, and the mass of their subhaloes depends on the resolution and on the number of particles chosen for a subhalo to be a genuine one (20 in our case). Nevertheless, when a subhalo is no longer recognized in the merger tree, either because disrupted or because its mass went under the resolution of the simulation, our SAM keeps following the satellite it was hosting. These satellites are the so-called *orphans*, and constitute the kind of satellites to which the model so far described is applied. The other population of satellites, those that keep maintaining a subhalo, are also subject to the same prescription, but in their case a further condition must be satisfied first:

$$R_{half}^{DM} < R_{half}^{Disk}, \qquad (2)$$

where $R_{half}^{DM}$ is the half-mass radius of the parent subhalo, and $R_{half}^{Disk}$ the half-mass radius of the galaxy's disk, that is $1.68 \cdot R_{sl}$ for an exponential profile. As mentioned earlier, some of the ICL in a halo can be associated to satellites. This happens for satellites that still have their own subhalo and for which the extra condition for stripping is not satisfied. In general, as long as that condition is not met, these satellites are allowed to be surrounded by the amount of ICL that they are carrying since they were central galaxies. Once the condition is met, they lose the ICL associated with them, which is transferred to that associated with the BCG.

The prescription just described accounts for one way to generate ICL during the growth of a cluster. The other important channel is given by mergers between the central galaxy and satellites, during the violent relaxation processes that take place in the merger. Practically speaking, this channel is very simple. At each merger, minor or major, we assume that part of the stellar mass of the merging satellite becomes unbound, and joins the ICL component of the corresponding central galaxy. In C14 this fraction was quantified to be 0.2 (see that paper for more details). Later works double-checked (Contini et al. 2018, 2019, 2022) the assumption of using the same fraction in larger simulations (with respect to that at which the result in C14 relied on), thus confirming to be a good approximation of the amount of stellar mass that gets unbound during a merger [4].

The stellar stripping and merger channels are the *direct* ways to produce ICL in our model. Another possible process would be *in-situ* star formation, but it has been discarded by observations (e.g., Melnick et al. 2012) and so never been implemented in the model. However, there is also an *indirect* way for BCGs to accumulate ICL, and it is given by the accretion of pre-processed ICL during mergers (minor and major) between haloes. This amount of ICL is then produced elsewhere, i.e. in a group from the two direct channels, and later accreted by another halo. The accreted ICL is essentially provided by satellite galaxies in two possible ways:

a) accretion of ICL associated to satellites with their own subhalo which experience the first stripping event;

b) accretion of ICL associated to satellite galaxies that become orphans (i.e. they have lost their subhalo).

In C14 we have shown that the pre-processing channel can be very important, in particular on cluster scales ($\log M_{halo} > 14.5$). It is not a goal of this paper to quantify the accretion history of the ICL and neither the importance of stripping/mergers as a function of time (which will be the main goal of a forthcoming study), but it is worth pointing out that, once in-situ star formation is ruled out as a possible channel, *the ICL forms only through stripping and mergers*, while pre-processing and accretion remain an indirect way for clusters to increase, through already formed ICL, their own component. Under this point of view, pre-processing can be considered a sub-channel.

Before concluding this section, there are other two important modifications with respect to the original model (C14) that are worth mentioning. Similarly to the way the SAM is able to compute magnitudes in several bands for galaxies, it does also for the ICL. This allows us to probe more observed properties of the ICL, such as luminosities and colors (see, e.g., Contini et al. 2019, 2022). The second, also implemented for the analysis done in Contini et al. (2019), relies on the metallicity of the ICL. While in C14 the assumption was that at a given fraction of mass stripped and moved to the ICL component the same fraction of metals were moved, in Contini et al. (2019) the model assumes a metallicity gradient in satellite galaxies. In practice, at each episode of stripping, metals are moved from the satellite by assuming that the scale length of the metals $R_{sl,metals}$ is a fraction $f_R$ of the total stellar mass in the disk, $R_{sl,metals} = f_R R_{sl}$, being $f_R$ randomly chosen between 0.5 and 1 (in order to assume a more concentrated distribution of metals than stars) for each satellite (see Contini et al. 2019 regarding the implications of this assumption).

---

[4] However, given the wide distribution of fractions found, the scatter in the distribution itself might be due to different properties of the satellite and on the circularity of its orbit. This still leaves room for a further development of this channel.

## 3. RESULTS

As briefly anticipated in Section 1, the main goal of this work is to find a link between the formation of the ICL and the dynamical state of the host group/cluster. From a theoretical point of view, if stellar stripping and mergers are the two processes responsible for the formation of most of the ICL, then there must be a correlation between the growth stage of a halo and the amount, or the fraction, of ICL in it. In the analysis below we are then going to explore possible correlations between the fraction of ICL, quantified as $f_{ICL} = M_{ICL}/(M_{ICL} + M_{BCG} + M_{SAT})$, where the quoted stellar masses are those of the ICL (which includes the stellar halo because it is a component not explicitly implemented in the SAM), of the BCG, and the total stellar mass of the satellite galaxies within the virial radius $R_{200}$, and a few halo properties that can be used as tracers of their dynamical state, such as the virial mass $M_{200}$, the concentration $c_{200}$, and the fraction of early-type galaxies within $R_{200}$.

Before going through the analysis, we want to point out that the results shown below are not dependent either on the refinement of the data storage of the runs, or on the range of mass resolutions spanned by the set of simulations. We have tested the relevant trends in the analysis below by comparing YS200 (lowest resolution) alone, with YS125 and YS50 (highest resolution) together, finding no significant deviations, in particular on the ICL fraction as a function of halo mass (Figure 2), which has been found in C14 to be sensitive to the mass resolution.

### 3.1. ICL Fraction and Halo Mass

We start our analysis in Figure 1, by showing a simple distribution of the fraction of ICL in our sample of haloes. Here, and in the rest of the analysis, with the only exception of Figure 6, our sample comprises 2509 haloes at redshift $z = 0$ with mass larger than $\log M_{halo} \sim 13$ and up to $\log M_{halo} \sim 15$, containing at least 4 satellites with $\log M_* \geq 8.5$. The solid black line represents the prediction of our model, while the dashed red line is the resulting gaussian fit (using IDL GAUSS-FIT with nterms= 5). The fraction of ICL covers a wide range, from almost zero to around 0.8. The distribution peaks at $f_{ICL} \sim 0.35$ with a standard deviation of about 0.12, in good agreement with observational results in the local universe which collocate the fraction of ICL in groups and clusters in a range between 5% and 50% (see the reviews by Contini 2021 and Montes 2022).

The interesting message of the figure, which has never been shown so far, is that the fraction of ICL can be approximated with a normal distribution. Hence, this translates in a stochastic nature of those processes that are responsible for the formation (and evolution) of the ICL, such as stellar stripping, mergers and accretion of pre-processed unbound stars.

The distribution shown in Figure 1 is wide enough to cover the observed fractions, but two important points still remains unclear: (1) Does $f_{ICL}$ correlate with the mass of the host?; (2) What is the origin of the (wide) scatter? In C14 we proved that a substantial part of it in haloes of similar mass can be attributed to the concentration, and we did not find a clear correlation between $f_{ICL}$ and halo mass. Here, given the larger samples of haloes (around a factor of 7.5) with respect to that used in C14, we want to re-address the two points.

Figure 2 shows the fraction of ICL versus the mass of the host halo, color coded with respect to the concentration of the halo. The concentration $c_{200}$ has been computed with the assumption of an NFW (Navarro et al. 1997) profile, and defined as the ratio between the virial and the scale radii, i.e. as:

$$c_{200} = \frac{R_{200}}{R_S},$$

being the scale radius the distance from the centre where the slope of the profile approximately changes from -2 to -3. The concentration is an important parameter that gives an idea of how relaxed a halo is, and so about its dynamical state: at fixed mass, low values of $c_{200}$ indicate objects that are still on the way of their formation, while large values of $c_{200}$ belong to haloes that are on the final stage of their growth, or already relaxed. Circles from blue to red indicate an increasing concentration. The region between the two dotted black lines represents the $\pm 1\sigma$ distribution of the model predictions, while the solid black line represents the mean in each bin. It is clear from the plot, by looking at the distribution, that $f_{ICL}$ does depend (although very weakly) on the halo mass, up to groups with mass $\log M_{halo} \sim 14$, and constant on 0.4 up to clusters. In C14 we did not find a clear correlation in the analogous plot, rather a flat relation with a large scatter on group scales (where we had a larger statistics). Another important trend shown by this plot is that more concentrated haloes tend to populate the upper part of the distribution and those less concentrated the bottom part. This translates, on average, in a higher fraction of ICL for more concentrated objects. The trend is possible to be seen by comparing the solid red line with the blue one, which represent the average ICL fractions for the most and least concentrated haloes, respectively. The full picture coming from

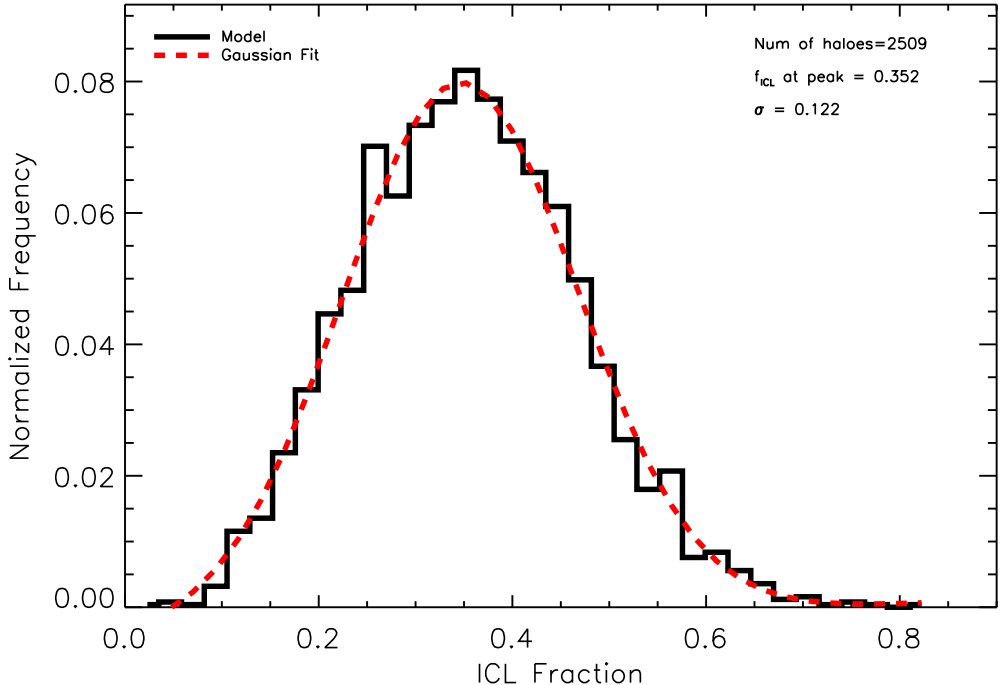

**Figure 1.** Distribution of the fraction of the ICL in our sample of haloes (solid black line) and the corresponding Gaussian fit (IDL GAUSSFIT with nterms= 5, dashed red line). Here (and below unless otherwise specified) we sample only haloes with $\log M_{halo}[M_\odot/h] \geq 13.0$ containing at least 5 galaxies (including the central one) with $\log M_* \geq 8.5$. The total number of haloes in our sample is 2509 from $\log M_\odot[M_\odot/h] \geq 13.0$ up to more than $\log M_{halo}[M_\odot/h] \geq 15.0$, i.e. spanning a wide range of halo masses. The distribution peaks on $f_{ICL} \sim 0.35$, and the error is modest, $\sigma \sim 0.12$, which put together give rise to a medium wide distribution of ICL fraction as predicted by several observations and numerical methods (see, e.g., the reviews by Contini 2021; Montes 2022). The fact that the distribution of the ICL fraction in galaxy groups and clusters can be approximated with a gaussian also highlights the stochastic nature of the processes that bring to the formation of the ICL, such as stellar stripping and galaxy mergers.

Figure 2 can be read as follows: $f_{ICL}$ slowly increases as the halo mass increases but, at the same time, more concentrated haloes, regardless of their mass, tend to host a higher fraction of ICL. In the light of these results, the concentration seems to have a major role in shaping the relation shown in Figure 2, scatter included. We will come back on this point in the following section.

### 3.2. ICL Fraction and Concentration

In Figure 3 we directly investigate on the relation between $f_{ICL}$ and $c_{200}$. The range in concentration has been divided in 10 bins all containing an equal number of haloes, and their ICL fraction averaged (purple diamonds) on each bin of concentration. The solid black line represents the linear fit ($\chi_r^2 = 1.04$) done on the 10 data points. Differently from Figure 2, in this plot the colors of the circles represent the number of galaxies within the virial radius, as marked in the legend, i.e. an increasing number from blue to red. Colors tending to red stand for objects more and more massive. It is clear from the fit (solid black line), but also from the data distribution, that $f_{ICL}$ increases with increasing concentration. The fraction of ICL increases from about

0.25 at low concentrations, to about 0.45 at high concentrations (confirming the trend shown by the colors of the circles in Figure 2), which is remarkable considering that about 70% of the ICL fractions are located between 0.23 and 0.47 ($\pm 1\sigma$ region in Figure 1). The result can be explained by the fact that more concentrated haloes are dynamical older, in a more advanced stage of their growth. As such, also the process of forming ICL is in advanced stage, since in more concentrated haloes the probability of stellar stripping is higher.

Higher concentrations means, on average, higher ICL fractions, in the sense that the concentration can explain part of the scatter seen in Figure 2. A question arises naturally: how much of this scatter the concentration can effectively explain? Or, put in other words: what is the efficacy (not efficiency) of the concentration in shaping the scatter such that the higher the concentration, the higher $f_{ICL}$? We want, somehow, to quantify the efficacy of the concentration and, to do that, we introduce two quantities. One quantity, $\eta^+$, which quantifies the percentage of haloes that have both ICL fraction and concentration above their respective averages, and the quantity $\eta^-$, which quantifies the percentages of haloes that have both ICL fraction and concentration below their respective averages. Mathematically, the two efficacies are defined as follows:

$$\eta^+ = \frac{N_{haloes}(c_{200} > \bar{c}_{200} \wedge f_{ICL} > \bar{f}_{ICL})}{N_{TOT}};$$

$$\eta^- = \frac{N_{haloes}(c_{200} < \bar{c}_{200} \wedge f_{ICL} < \bar{f}_{ICL})}{N_{TOT}},$$

and the total efficacy $\eta$ is given by the sum of $\eta^+$ and $\eta^-$. An $\eta$ close to 1 indicates a more efficacious concentration.

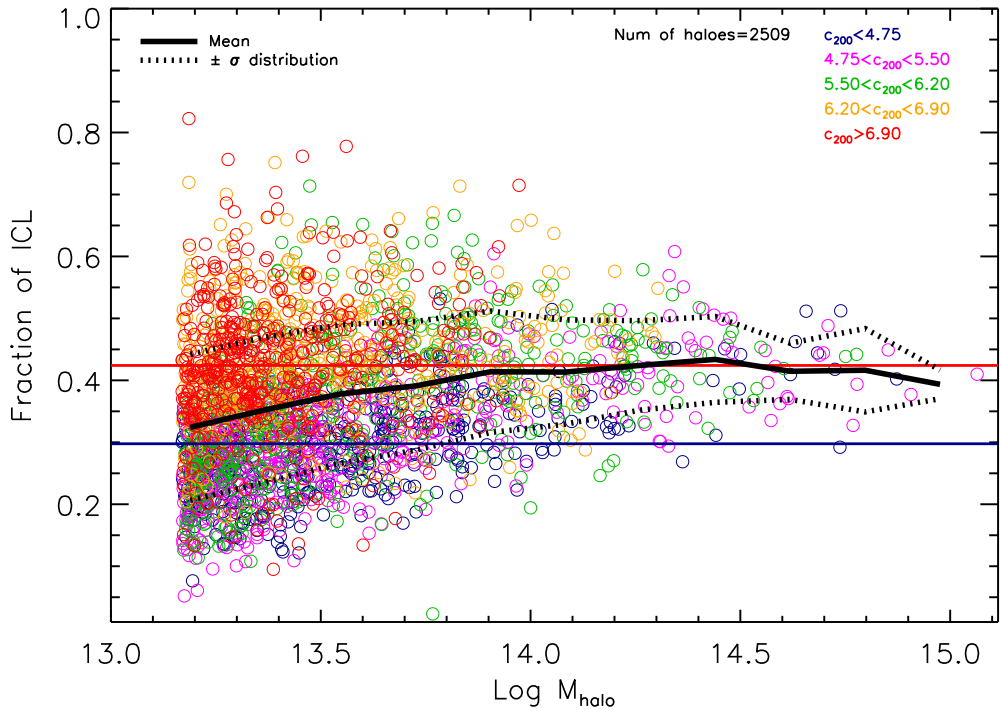

**Figure 2.** Fraction of ICL as a function of halo mass for haloes with different concentration $c_{200}$ (circles with different colors) derived by assuming an NFW profile (see text for more details). The region between the two dotted lines shows the $\pm 1\sigma$ distribution, while the solid black line is the mean of the data. The solid red and blue lines represent the average ICL fractions in the lowest and highest ranges in concentration, respectively. Contrary to what found in C14 where the sample of haloes was much smaller, here we do find a positive (although weak) trend between the fraction of ICL and its host halo mass, which is clear from the distribution up to haloes with $\log M_{halo} \sim 14.0$, and then it stays constant at $\sim 0.4$. Another important result coming from the plot is the tendency of the ICL fraction to be gradually higher for more concentrated haloes (colors from blue to red, from the bottom to the top of the figure). Moreover, the most massive haloes are the least concentrated ones, as they are supposed to, given the well-known relation between halo mass and concentration (e.g., Prada et al. 2012). The net message of the plot is that the ICL fraction weakly increases towards higher halo masses and, at a given halo mass, it is higher in more concentrated ones.

In Figure 4 we plot the total efficacy $\eta$ (blue), and the two partial $\eta^+$ (red) and $\eta^-$ (green) in five different halo mass bins. Translated in percentages, the total efficacy of the concentration is higher, $\eta \sim 0.8$, in the lowest halo mass bin (where we have the largest statistics), and decreases to $\eta \sim 0.5$ in the highest halo mass bin (where we have the lowest statistics). This means that only $\sim 20\%$ ($\sim 50\%$) of the least massive haloes (most massive haloes) have the concentration and the

ICL fraction that are outsiders in the trend seen in Figure 3. If we consider the mean concentration and the mean ICL fraction of the full sample rather than binning in 5 subsamples, the average efficacy is 73%. This percentage indicates that, overall, only 27% of the haloes in our sample have ICL fraction and concentration that are outsiders in the trend shown in Figure 3. In a picture where the concentration has to be higher/lower where the ICL fraction is higher/lower, our model predicts that the concentration is playing a substantial role in 73% of the cases.

### 3.3. ICL and Early-Type Galaxies Fractions

Another "tracer" of the dynamical state of haloes is the fraction of early-type galaxies (ETGs) within the virial radius, in the sense that we would expect a higher fraction of ETGs in dynamically older/more evolved objects (and so more concentrated). Given the correlations seen above, it might also be true that haloes which host a higher fraction of ICL should also host a higher fraction of ETGs. This relation has been recently investigated, observationally, by Ragusa et al. (2023), who found a very weak trend between the two properties. We will fully discuss the interpretation of their results in the following section, when we compare the messages given by our analysis with their conclusion. In Figure 5 we plot the equivalent of their Figure 3, i.e. the fraction of ICL as a function of the fraction of ETGs, defined here as the ratio between galaxies for which the bulge-to-total ratio (B/T) is larger than 0.7 [5], and the total number of galaxies within the virial radius. A direct comparison with their observed data is not possible, considering that the fraction of ETGs are defined differently (caveats will be addressed in Section 4), and our threshold in stellar mass of galaxies is much lower than theirs (we keep our threshold for consistency with the rest of the plots, and due to the fact that we are not doing a quantitative comparison). However, contrary to them, we do not find any correlation. We do find some trend, although very weak but close to their slope, if we increase at least one order of magnitude the threshold in stellar mass, from $\log M_* = 8.5$ to $\log M_* \gtrsim 9.5$ (not shown). One more information we can get from Figure 5 is that the concentration increases only vertically, towards higher ICL fractions, but there is no horizontal trend. This means that there is no clear correlation between the fraction of ETGs and the concentration of their host halo.

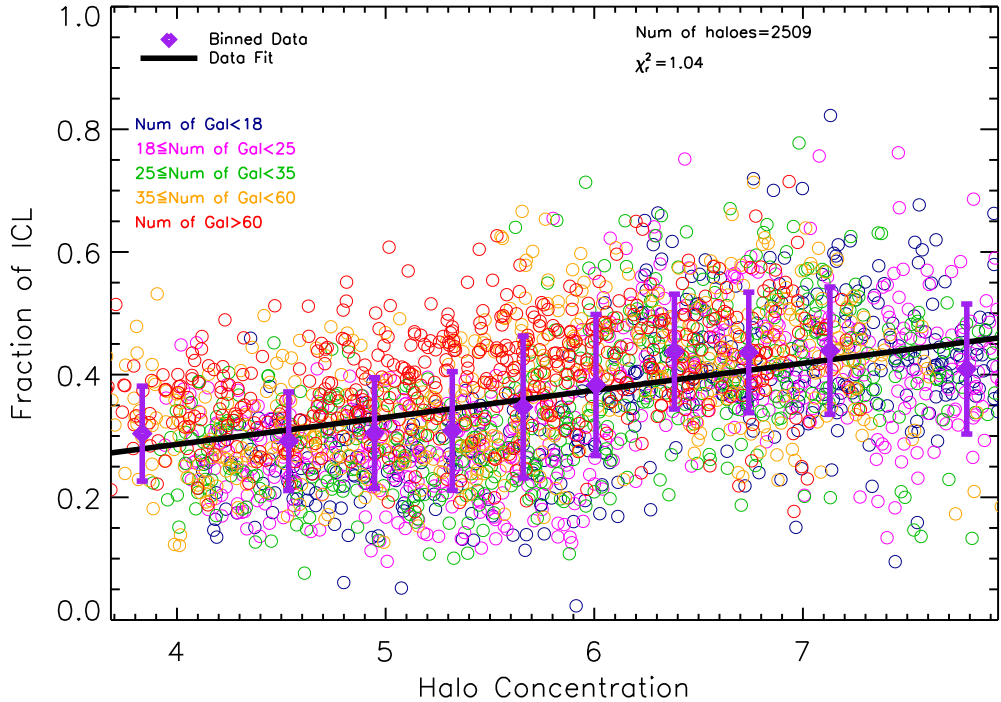

**Figure 3.** Fraction of ICL as a function of halo concentration $c_{200}$ (circles with different colors roughly indicate the number of galaxies within the virial radius). The range in concentration has been split in 10 bins all containing the same number of haloes, and their ICL fraction averaged (purple diamonds), while the solid black line represents their linear fit ($f_{ICL} = (0.044 \pm 0.008)c_{200} - 0.110 \pm 0.047$). As already pointed out in C14, the halo concentration can explain large part of the scatter in the relation between the ICL fraction and the mass of the host halo. Here, with a much larger number of haloes we show that $f_{ICL}$ does depend on the concentration of the haloes (as anticipated in Figure 2), although the dependence is seen more clearly in the fit rather than the data, which instead show a larger scatter. More concentrated haloes can be translated in dynamically older objects, for which the process of formation of the ICL is in an advanced stage, given the fact that in a more concentrated halo the probability of stellar stripping is higher.

Considering the trends seen in the former figures, the fraction of ICL tends to be gradually higher in more massive haloes, and the scatter of this relation can be explained by the range of concentrations in haloes of similar mass (given the scatter in the relation between halo mass and concentration). Not only, more concentrated haloes tend to have a higher fraction of ICL, which can lead to the statement that, on average, dynamically older or more evolved objects host a higher fraction of ICL. The prediction of our model is the net result of two distinct effects: (a) in more concentrated haloes the probability of stellar stripping is higher and, (b) stellar stripping acts first on more fragile galaxies such as the disk of spirals (Contini et al. 2018), thus favoring the increase in the fraction of early-type galaxies. Nevertheless, being the concentration a parameter that acts in the innermost regions of the haloes, an increasing fraction of ETGs with increasing concentration is not seen, and this reflects also on the dependence of ETGs and ICL fractions.

---

[5] The bulge-to-total ratio B/T is the only possible way to define the morphology of a galaxy in SAMs. Our B/T> 0.7 for early-type galaxies is the most common value frequently used in observational studies (e.g., Oh et al. 2018).

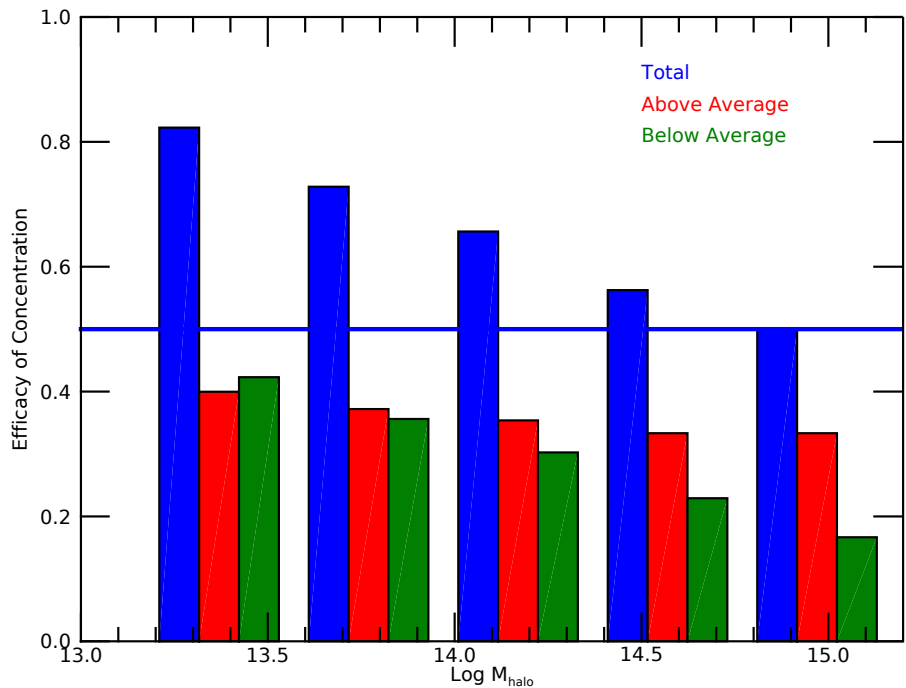

**Figure 4.** In each halo mass bin, fraction of haloes having concentration and fraction of ICL above (red) and below (green) their averages in the bin at the same time. The blue color refers to the sum of red and green, and the horizontal blue line marks the medium efficacy 0.5. As argued in the text, it is a way to quantify whether or not concentration and fraction of ICL go in the same direction if the former is driving the bulk of ICL formation through stellar stripping, i.e. higher fractions of ICL in haloes more concentrated. At least 60% of haloes in every halo mass bin, except the most massive ones where we have only six objects, have either higher ICL fractions and concentrations above their mean, or below, and the other 40% with the two quantities going in opposite directions. The percentages increase in lower mass haloes, by reaching ∼ 80% in the lowest halo mass bin, which is also the one containing the largest statistics.

Under this point of view, if stellar stripping (but also mergers fit in the scenario) is the main channel for the formation of the ICL, the concentration $c_{200}$ is the main driver of the trends seen above and so, it is also driving the formation of the ICL. We will come back on this point below, given the fact that it is the main goal of this work, by discussing the implications of our results in a context where stellar stripping is not yet believed to be as important as highlighted here (and in former works).

### 3.4. ICL Luminosity and Halo Mass

Before concluding this section, as a comparison between our model with very recent observations, we want to show in Figure 6 the luminosity of the ICL in g-band, $L_{ICL}$, as a function of the halo mass. Our model predictions (circles with different colors) are compared with 17 observed groups and cluster (purple diamonds) from Ragusa et al. (2023) (VEGAS survey, Capaccioli et al. 2015; Iodice et al. 2021), and 14 groups/clusters (cyan triangles) from the sample of Kluge et al. (2021). On top of them, we also plot the fit ($\chi_r^2 = 1.18$) of our data (solid black line) and the fit including all the observed data together (purple line). It must be noted that for this figure, in order to cover the range in halo mass spanned by the observed data, we make use of the HR simulations and reach halo masses down to log $M_{halo} \sim 12.6$, without any possible drawback given by the resolution of the simulations. For this reason, the number of haloes in the new sample increases from about 2500 to more than 7700. Our predictions compare fairly well with the observed data in a large part of the halo mass range probed. Indeed, most of the observed data points lie in the cloud of our data, and only the least and most massive objects are "outsiders" (all the others lie within $\pm 2\sigma$ of our distribution). However, by comparing the fit of our data with that of the observed ones, we notice that they have quite different slopes (and intercepts), which means that the two distributions are not statistically close. If we remove from the fit of the observed data the haloes that do not lie within $\pm 2\sigma$ of our distribution, the new fit (not shown) would be much closer to that of our model predictions. Hence, the fit of the observed data is sensitive to the number of objects due to the fact that they span a wide range in luminosity.

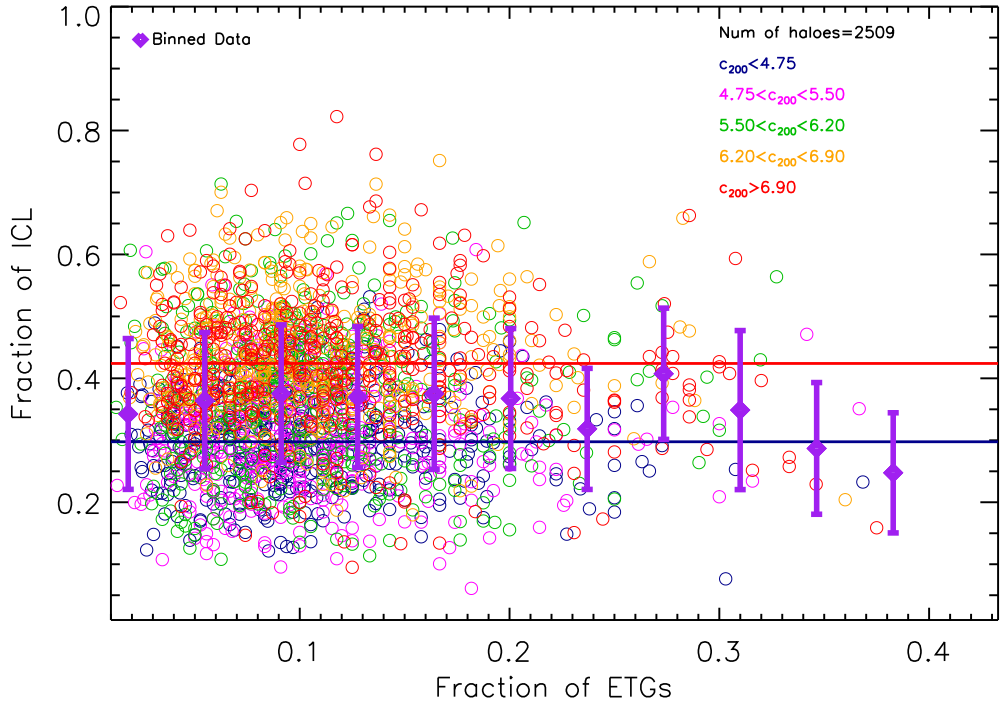

**Figure 5.** Fraction of ICL as a function of the fraction of early-type galaxies (ETGs) within the virial radius $R_{200}$, for the same sample of haloes considered in the former figures. ETGs are defined as those galaxies for which the bulge-to-total ratio (B/T) is larger than 0.7 (see text for more details). As in Figure 2, colors from blue to red indicate an increasing concentration, while purple diamonds the mean ICL fractions in bins of ETGs fractions. The solid red and blue lines represent the average ICL fractions in the lowest and highest ranges in concentration, respectively. We do not find any correlation unless we increase the threshold in stellar mass by an order of magnitude, from $\log M_* = 8.5$ to $\log M_* = 9.5$ (not shown). This result somehow echoes that found by Ragusa et al. (2023), who show that the observed fraction of ICL in the VEGAS survey is slightly higher in groups/clusters having a larger fraction of ETGs. However, a direct comparison with their data is not possible due to the different definitions of ETGs used (see the text for a full discussion about this point).

In the following section we discuss deeper the results of our analysis and their implications, mainly by focusing on the role that the halo concentration plays in the formation of the ICL, together with some caveats worth mentioning and making plain.

## 4. DISCUSSION

The idea of using the ICL as a tracer of the dynamical state of their host haloes has been having a fast development in the recent few years (Zibetti et al. 2005; Giallongo et al. 2015; Montes & Trujillo 2019; Alonso Asensio et al. 2020; Contini & Gu 2020; Ragusa et al. 2023 and others). An example of the importance of the ICL lies in utilizing it as a luminous tracer of the dark matter distribution (Montes & Trujillo 2019) since, by definition, it follows the potential well of the host haloes, and then its distribution is expected to be very close to the dark matter one (Zibetti et al. 2005; Montes & Trujillo 2019; Contini & Gu 2020, 2021). The literature has plenty of examples of studies that followed this direction of investigation, both observational and theoretical (see references above and therein). For example, Contini & Gu (2020) developed a simple model to describe the distribution of the ICL by linking it to the dark matter profile. The key-point of the model is the assumption that the ICL concentration is strictly linked with the halo concentration. Taking profit of observational results, the model was tuned to find the link between the two concentrations and later used to build a full BCG+ICL profile, which led to find the typical transi-tion radius between the region dominated by the BCG, and that dominated by the ICL (Contini et al. 2022).

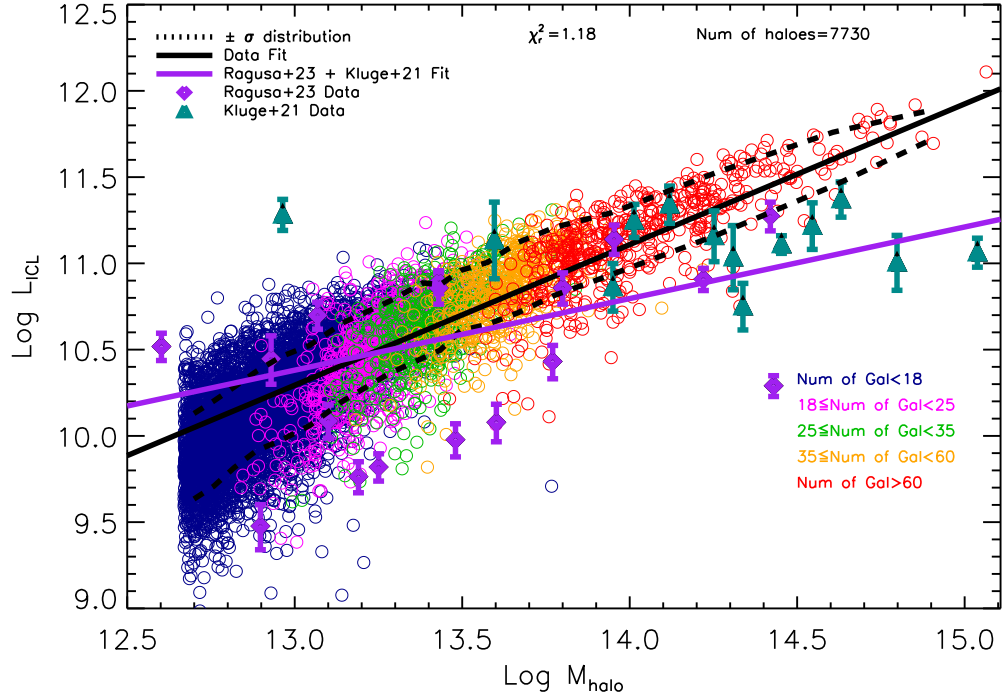

**Figure 6.** The luminosity in g-band of the ICL as a function of halo mass (circles with different colors according to the number of galaxies within the virial radius, and their fit ($\log L_{ICL} = (0.82 \pm 0.02) \log M_{halo} - 0.31 \pm 0.27$) represented by the solid black line), compared with the observed ones by Ragusa et al. (2023) (purple diamonds) and by Kluge et al. (2021) (cyan triangles), and their fit including all of them (solid purple line). For a full coverage of the halo mass range spanned by the observed data, in this figure we take advantage of the high-resolution runs (see Table 1), then increasing the sample of haloes from $\sim 2500$ to more than 7700. Our model predictions compare fairly well with the observed data in a large part of the range explored, considering that most of the observed data points lie in the cloud of the predictions of our model, although it must be pointed out that the slopes of the two global relations (fits) are remarkably different. We refer the reader to the text for a full discussion of this point.

It appears then clear that the concentration must have a key role in the formation of the ICL. Before going into the detail of the analysis, we would like to point out how the stochastic nature of the hierarchical clustering reflects also on the formation of the ICL. In fact, dark matter haloes grow in time via mergers, minor or major, that are purely stochastic events. These building blocks for the final relaxed haloes bring also the material for the ICL. On a first level, they not only bring the pre-processed ICL, but more importantly, galaxies that will be subject to stellar stripping and mergers with the BCGs. The normal distribution of the ICL fractions seen in Figure 1 is then the reflection of the stochastic nature of the build-up of structures, with small haloes merging together to form larger ones and, at the same time, growing in stellar mass by the accretion of the stellar material.

The fundamental question is: what does really happen within the haloes when stars get unbound to form the ICL? As discussed previously, mergers and stellar stripping are the processes that lead to the formation of the ICL. Mergers with BCGs are the inevitable fate of satellites, a fate that is controlled by dynamical friction (Chandrasekhar 1943). Satellite galaxies orbiting within the halo feel a drag force caused by the surrounding material, which has the net result of slowing them down and making them approach the centre faster. Dynamical friction is directly proportional to the mass of the satellite, which means that more massive satellites have a shorter dynamical friction time (that translates in a shorter merging time) and end up in the innermost regions of the cluster faster than less massive satellites (Contini & Kang 2015; Kim et al. 2020, and references therein). Here it comes the role of the concentration in driving the formation of the ICL through stellar stripping. Indeed, the innermost regions are also the most concentrated ones, where tidal forces are stronger. More massive satellites get faster in these regions, are more subject to tidal forces and contribute remarkably to enhance the ICL component.

If the halo mass is driving the formation of the ICL, we would not expect any strong correlation between the ICL fraction and the halo mass itself, given the fact that more massive haloes would host a larger amount of ICL, but also a larger amount of stellar mass, so that the proportional contribution given by the ICL would be constant or weakly dependent on the halo mass. What really makes the difference is the fact that, although there is a correlation between mass and concentration in clusters, that relation shows some scatter. This basically means that haloes of the same mass can show a range of concentrations, and depending on how concentrated they are, i.e. more or less evolved, the fraction of ICL within them can be different. It is exactly the message that Figure 2 is showing, where the fraction of ICL slowly increases from small to large groups. The trend up to $\log M_{halo} \sim 14$ followed by a constant fraction until cluster scales is clear. Not only, but looking at the colors of the circles, which indicate the ranges of concentration, there is the tendency for more concentrated haloes to host a higher fraction of ICL, which is an indication that the concentration, rather than the mass, is driving the content of ICL through stellar stripping.

This is even more evident in Figure 3, where the direct relation between the fraction of ICL and concentration is shown. The distribution of the data is very scattered, but a positive trend is visible. A linear fit provides a shallow slope, consistent with zero within $6\sigma$. Moreover, it must be noted that 70% of the ICL fractions in the distribution shown in Figure 1 lie in the range 0.23-0.47 ($\pm 1\sigma$ region of the distribution), which is roughly the range in ICL fractions spanned by their relation with the concentration (Figure 3). In C14 we did not find a clear trend between the ICL fraction and the halo mass, but we noticed that the concentration was driving the large scatter in the relation. The caveat is that, while in C14 we had a sample of slightly more than 300 haloes for the same range of halo mass probed here, now we account for more than 2500 haloes. The enlarged sample populates mostly small and large groups, but, as already pointed out, still lacks enough statistics on very massive clusters (with mass $\log M_{halo} > 14.7$). However, increasing the number of objects on small and intermediate scales allows us to finally answer to the question: does the ICL fraction depend on the halo mass? The answer given by our model is yes, the ICL fraction does depend, but very weakly, on the halo mass up to group scales ($\log M_{halo} \sim 14$) to flatten to a constant value afterwords, and the scatter seen in the relation is driven by the concentration. That is not all. The concentration correlates directly with the ICL fraction. This means that the concentration is a more suitable candidate for driving the formation of the ICL, at least up to large group size haloes. A quantification of the role played by the concentration in explaining the scatter seen in Figure 2, i.e. higher/lower ICL fraction in haloes with higher/lower concentration has been given in Figure 4. Depending on the halo mass bin, the mean efficacy ranges from 0.5 (cluster scale) to 0.8 (small group scale), with an average efficacy of 0.73. This means that only in 27% of our haloes, the concentration alone cannot account for, or over explains the ICL fraction found. In the first case (concentration below and ICL fraction above the averages), some other mechanisms act with a similar strength, e.g. more material from pre-processing or from mergers. In the second case (concentration above and ICL fraction below the averages), the most likely explanation is given by a different number of intermediate/massive satellites subject to stripping during the halo assembly. Indeed, as shown by Purcell et al. (2007) and further confirmed by C14, the ICL fraction depends strongly on the number of survived satellites.

The halo concentration, however, is not a direct observable. As suggested recently by Ragusa et al. (2023) (but see also Da Rocha et al. 2008; Ragusa et al. 2022 and references therein), the fraction of ETGs can be an observable tracer of the dynamical state of haloes. Objects that have a larger fraction of ETGs have been shown to have also a larger fraction of ICL. The picture is as follows: the fraction of ETGs is higher in more evolved haloes, which in turn are older and had more time to form (or accrete) their ICL. Ragusa et al. (2023) use this argument to link the ICL and ETGs fractions, so to find an observable tracer that could be the driver of the ICL formation. Shortly, they investigate the relation between the ICL fraction and halo mass, finding no clear trend. However, when they plot the fraction of ICL against the fraction of ETGs, they do find a weak correlation. Their claim is that the halo mass is driving the formation of the ICL if it first forms on group scales, and later it is accreted on cluster scales. According to them, this would explain both the lack of a strong correlation between ICL fraction and halo mass, and the mild correlation between the ICL fraction with the ETGs one at the same time.

We have investigated the relation between the fractional budget of ICL and that of ETGs, color coded with the halo concentration, in Figure 5, in order to examine the claim by Ragusa et al. (2023). As discussed in Section 3.3, we do not find a correlation between the two fractions. A direct quantitative comparison with their data is not possible, given the fact that their definition of ETGs is not the same as ours. Indeed, while they use the classification from the NASA/IPAC Extragalac-tic Database (NED) [6], and classify some of them (not present in NED) according to their photometric properties including B/T (private conversation), our classification entirely relies on their B/T (larger than 0.7). However, although not shown for consistency with the other plots, when we increase the threshold in stellar mass by an order of magnitude in sampling satellites, we do find a weak correlation similar to theirs in terms of slope. The main difference between our and their analysis consists in the fact that, while we do find a weak correlation between the fraction of ICL and the halo mass up to $\log M_{halo} \sim 14$, they do not. As discussed above, we find a correlation between the ICL fraction and the halo concentration, which makes a more physical sense if stellar stripping is driving the bulk of the ICL formation.

A caveat that must be discussed concerns the difference in the samples. First and foremost, their sample extends up to halo mass $\log M_{halo} \sim 14.5$ and do not reach masses over $\log M_{halo} \sim 15$ as our does. Second, their sample is uniformly distributed over the range of halo mass, but with a much lower number of objects. Nevertheless, we do find similar amount of ICL as shown by the comparison of the ICL luminosity in Figure 6. Considering the scatter in the ICL fraction versus halo mass relation (or concentration), it is likely that a much larger sample of observed objects with mass up to that of large groups and clusters would bring to a correlation similar to what we find, and predicted by other theoretical models (Lin & Mohr 2004; Murante et al. 2007; Purcell et al. 2007; Henden et al. 2019). A similar correlation would be, in principle, possible to be seen even in observations, by collecting all the measurements from different authors (in the local universe). However, the fraction of ICL is strongly dependent on the radius within which it is calculated (Montes & Trujillo 2018; Contini 2021; Montes 2022), and considering that different authors often used different apertures for measuring the ICL, such a global collection would be biased.

To conclude, given the results shown in the analysis and the above discussion of them, we suggest that the halo concentration is driving the bulk of the formation of the ICL. Pre-processing/accretion is just a sub-channel for accumulating more ICL that was formed anyway through stellar stripping and mergers. The latter have been shown to contribute up to $\sim 20\%$, which leaves the rest to tidal stripping and disruption (probed to be almost negligible in terms of mass deposited in the ICL). The global picture sees part of the ICL form-ing in group scales via stellar stripping and mergers, and later accreted on cluster scale. At the same time in clusters (but also in smaller objects), the formation of ICL keeps going through mergers, but mostly through stellar stripping. What makes the difference in the amount, or fractional budget, of the ICL, is how concentrated a halo can be. In more concentrated objects, tidal forces are stronger and responsible for pushing up the amount of stars that get unbound and join the diffuse component and, at the same time, for reducing the budget of bound stars. On cluster scales the concentration is low, but these objects can count on a much larger number of intermediate/massive satellites (potentially subject to stellar stripping) than lower mass haloes. With a much larger statistics on cluster scales ($\log M_{halo} > 14.7$), it would be very likely to see the same trend ICL fraction-concentration seen in smaller objects.

## 5. CONCLUSIONS

Taking advantage of a set of dark matter only simulations on mergers trees of which we run our state-of-art semi-analytic model, we investigated the role that the halo concentration plays in the formation of the ICL. We used a wide sample of haloes, from small group scale to large clusters, to analyze the correlations between the fraction of ICL, halo mass/concentration, and the fraction of early-type galaxies within the virial radius. On the basis of our analysis and discussion of the results shown, our main conclusions are as follows:

- The fraction of ICL in the range of halo mass probed is normally distributed, a result that derives from the stochastic nature of those processes responsible for the formation of the ICL, i.e. stellar stripping and mergers;

- By increasing the number of objects in the sample, with respect to previous works, we find that the ICL fraction weakly depends on halo mass from small to intermediate groups ($\log M_{halo} \sim 14$), then converges to a constant value, around 0.4, up to clusters with $\log M_{halo} \sim 15$. However, it must be noted that, in general, our results are statistically safe up to large group scales, $\log M_\odot \sim 14.7$;

- The fraction of ICL increases as a function of halo concentration and, at the same time, it does not depend on the fraction of early-type galaxies, as found in recent observations. More concentrated haloes are dynamically older, in a more advanced stage of their growth. This explains the possibility that the fraction of ICL is higher in haloes dynamically older, because in more concentrated haloes stellar stripping is more likely, leading to an increase in the amount of diffuse light and a decrease in that of bound stars at the same time;

- The halo concentration is probably driving the bulk of the ICL formation. The ICL forms through stellar stripping of intermediate/massive satellites that reach the innermost regions faster than the less massive ones. In these regions, where the concentration plays a remarkable role, tidal forces are stronger and the production of diffuse light is boosted.

In a forthcoming series of studies, we will extend the analysis done in this work to higher redshifts, by looking at how the correlations evolve as a function of time when the ICL is forming and then assembling. A particular attention will be given to the contributions provided by the different channels as a function of time, followed by an investigation of what happens on smaller scales, in haloes with $\log M_{halo} \leq 13$.

## ACKNOWLEDGEMENTS

The authors thank the anonymous referee for his/her constructive comments which helped to improve the manuscript, and Rossella Ragusa for providing the data and for useful information about them. E.C. and S.K.Y. acknowledge support from the Korean National Research Foundation (2020R1A2C3003769). E.C. and S.J. acknowledge support from the Korean National Research Foundation (RS-2023-00241934). All the authors are supported by the Korean National Research Foundation (2022R1A6A1A03053472). J.R. was supported by the KASI-Yonsei Postdoctoral Fellowship and was supported by the Korea Astronomy and Space Science Institute under the R&D program (Project No. 2023-1-830-00), supervised by the Ministry of Science and ICT.

## REFERENCES

Alonso Asensio, I., Dalla Vecchia, C., Bahé, Y. M., Barnes, D. J., & Kay, S. T. 2020, MNRAS, 494, 1859, doi: 10.1093/mnras/staa861

Annunziatella, M., Biviano, A., Mercurio, A., et al. 2014, A&A, 571, A80, doi: 10.1051/0004-6361/201424102

Annunziatella, M., Mercurio, A., Biviano, A., et al. 2016, A&A, 585, A160, doi: 10.1051/0004-6361/201527399

Arnaboldi, M., & Gerhard, O. 2022, Frontiers in Astronomy and Space Sciences, 9, 403, doi: 10.3389/fspas.2022.872283

Binney, J., & Tremaine, S. 2008, Galactic Dynamics: Second Edition

Burke, C., Hilton, M., & Collins, C. 2015, MNRAS, 449, 2353, doi: 10.1093/mnras/stv450

Capaccioli, M., Spavone, M., Grado, A., et al. 2015, A&A, 581, A10, doi: 10.1051/0004-6361/201526252

Chabrier, G. 2003, PASP, 115, 763, doi: 10.1086/376392

Chandrasekhar, S. 1943, ApJ, 97, 255, doi: 10.1086/144517

Conroy, C., Wechsler, R. H., & Kravtsov, A. V. 2007, ApJ, 668, 826, doi: 10.1086/521425

Contini, E. 2021, Galaxies, 9, 60, doi: 10.3390/galaxies9030060

Contini, E., Chen, H. Z., & Gu, Q. 2022, ApJ, 928, 99, doi: 10.3847/1538-4357/ac57c4

Contini, E., De Lucia, G., Villalobos, Á., & Borgani, S. 2014, MNRAS, 437, 3787, doi: 10.1093/mnras/stt2174

Contini, E., & Gu, Q. 2020, ApJ, 901, 128, doi: 10.3847/1538-4357/abb1aa

—. 2021, ApJ, 915, 106, doi: 10.3847/1538-4357/ac01e6

Contini, E., Gu, Q., Ge, X., et al. 2020, ApJ, 889, 156, doi: 10.3847/1538-4357/ab6730

Contini, E., & Kang, X. 2015, MNRAS, 453, L53, doi: 10.1093/mnrasl/slv103

Contini, E., Kang, X., Romeo, A. D., & Xia, Q. 2017a, ApJ, 837, 27, doi: 10.3847/1538-4357/aa5d16

Contini, E., Kang, X., Romeo, A. D., Xia, Q., & Yi, S. K. 2017b, ApJ, 849, 156, doi: 10.3847/1538-4357/aa93dd

Contini, E., Yi, S. K., & Kang, X. 2018, MNRAS, 479, 932, doi: 10.1093/mnras/sty1518

—. 2019, ApJ, 871, 24, doi: 10.3847/1538-4357/aaf41f

Cui, W., Murante, G., Monaco, P., et al. 2014, MNRAS, 437, 816, doi: 10.1093/mnras/stt1940

Da Rocha, C., Ziegler, B. L., & Mendes de Oliveira, C. 2008, MNRAS, 388, 1433, doi: 10.1111/j.1365-2966.2008.13500.x

Davies, L. J. M., Robotham, A. S. G., Lagos, C. d. P., et al. 2019, MNRAS, 483, 5444, doi: 10.1093/mnras/sty3393

DeMaio, T., Gonzalez, A. H., Zabludoff, A., Zaritsky, D., & Bradač, M. 2015, MNRAS, 448, 1162, doi: 10.1093/mnras/stv033

DeMaio, T., Gonzalez, A. H., Zabludoff, A., et al. 2018, MNRAS, 474, 3009, doi: 10.1093/mnras/stx2946

Dolag, K., Murante, G., & Borgani, S. 2010, MNRAS, 405, 1544, doi: 10.1111/j.1365-2966.2010.16583.x

Edwards, L. O. V., Alpert, H. S., Trierweiler, I. L., Abraham, T., & Beizer, V. G. 2016, MNRAS, 461, 230, doi: 10.1093/mnras/stw1314

Giallongo, E., Menci, N., Grazian, A., et al. 2015, ApJ, 813, 68, doi: 10.1088/0004-637X/813/1/68

—. 2014, ApJ, 781, 24, doi: 10.1088/0004-637X/781/1/24

Gonzalez, A. H., Sivanandam, S., Zabludoff, A. I., & Zaritsky, D. 2013, ApJ, 778, 14, doi: 10.1088/0004-637X/778/1/14

Groenewald, D. N., Skelton, R. E., Gilbank, D. G., & Loubser, S. I. 2017, MNRAS, 467, 4101, doi: 10.1093/mnras/stx340

Gu, M., Conroy, C., Law, D., et al. 2020, ApJ, 894, 32, doi: 10.3847/1538-4357/ab845c

Guo, Q., White, S., Boylan-Kolchin, M., et al. 2011, MNRAS, 413, 101, doi: 10.1111/j.1365-2966.2010.18114.x

Henden, N. A., Puchwein, E., & Sijacki, D. 2019, MNRAS, 489, 2439, doi: 10.1093/mnras/stz2301

Iodice, E., Spavone, M., Cantiello, M., et al. 2017, ApJ, 851, 75, doi: 10.3847/1538-4357/aa9b30

Iodice, E., Spavone, M., Cattapan, A., et al. 2020, A&A, 635, A3, doi: 10.1051/0004-6361/201936435

Iodice, E., Spavone, M., Capaccioli, M., et al. 2021, The Messenger, 183, 25, doi: 10.18727/0722-6691/5232

Jaffe, W. 1983, MNRAS, 202, 995, doi: 10.1093/mnras/202.4.995

Joo, H., & Jee, M. J. 2023, Nature, 613, 37, doi: 10.1038/s41586-022-05396-4

Kawinwanichakij, L., Papovich, C., Quadri, R. F., et al. 2017, ApJ, 847, 134, doi: 10.3847/1538-4357/aa8b75

Kim, S., Contini, E., Choi, H., et al. 2020, ApJ, 905, 12, doi: 10.3847/1538-4357/abbfa6

Kluge, M., Bender, R., Riffeser, A., et al. 2021, ApJS, 252, 27, doi: 10.3847/1538-4365/abcda6

Ko, J., & Jee, M. J. 2018, ApJ, 862, 95, doi: 10.3847/1538-4357/aacbda

Lin, Y.-T., & Mohr, J. J. 2004, ApJ, 617, 879, doi: 10.1086/425412

Longobardi, A., Arnaboldi, M., Gerhard, O., & Hanuschik, R. 2015, A&A, 579, A135, doi: 10.1051/0004-6361/201525773

Mao, Z., Kodama, T., Pérez-Martínez, J. M., et al. 2022, A&A, 666, A141, doi: 10.1051/0004-6361/202243733

Melnick, J., Giraud, E., Toledo, I., Selman, F., & Quintana, H. 2012, MNRAS, 427, 850, doi: 10.1111/j.1365-2966.2012.21924.x

Mihos, J. C., Harding, P., Feldmeier, J., & Morrison, H. 2005, ApJL, 631, L41, doi: 10.1086/497030

Mihos, J. C., Harding, P., Feldmeier, J. J., et al. 2017, ApJ, 834, 16, doi: 10.3847/1538-4357/834/1/16

Monaco, P., Murante, G., Borgani, S., & Fontanot, F. 2006, ApJL, 652, L89, doi: 10.1086/510236

Montes, M. 2022, Nature Astronomy, 6, 308, doi: 10.1038/s41550-022-01616-z

Montes, M., Brough, S., Owers, M. S., & Santucci, G. 2021, ApJ, 910, 45, doi: 10.3847/1538-4357/abddb6

Montes, M., & Trujillo, I. 2014, ApJ, 794, 137, doi: 10.1088/0004-637X/794/2/137

—. 2018, MNRAS, 474, 917, doi: 10.1093/mnras/stx2847

—. 2019, MNRAS, 482, 2838, doi: 10.1093/mnras/sty2858

Morishita, T., Abramson, L. E., Treu, T., et al. 2017, ApJ, 846, 139, doi: 10.3847/1538-4357/aa8403

Murante, G., Giovalli, M., Gerhard, O., et al. 2007, MNRAS, 377, 2, doi: 10.1111/j.1365-2966.2007.11568.x

Navarro, J. F., Frenk, C. S., & White, S. D. M. 1997, ApJ, 490, 493, doi: 10.1086/304888

Oh, S., Kim, K., Lee, J. H., et al. 2018, ApJS, 237, 14, doi: 10.3847/1538-4365/aacd47

Peng, Y.-j., Lilly, S. J., Renzini, A., & Carollo, M. 2012, ApJ, 757, 4, doi: 10.1088/0004-637X/757/1/4

Planck Collaboration, Aghanim, N., Akrami, Y., et al. 2020, A&A, 641, A6, doi: 10.1051/0004-6361/201833910

Prada, F., Klypin, A. A., Cuesta, A. J., Betancort-Rijo, J. E., & Primack, J. 2012, MNRAS, 423, 3018, doi: 10.1111/j.1365-2966.2012.21007.x

Puchwein, E., Springel, V., Sijacki, D., & Dolag, K. 2010, MNRAS, 406, 936, doi: 10.1111/j.1365-2966.2010.16786.x

Purcell, C. W., Bullock, J. S., & Zentner, A. R. 2007, ApJ, 666, 20, doi: 10.1086/519787

Ragusa, R., Mirabile, M., Spavone, M., et al. 2022, Frontiers in Astronomy and Space Sciences, 9, 852810, doi: 10.3389/fspas.2022.852810

Ragusa, R., Spavone, M., Iodice, E., et al. 2021, A&A, 651, A39, doi: 10.1051/0004-6361/202039921

Ragusa, R., Iodice, E., Spavone, M., et al. 2023, A&A, 670, L20, doi: 10.1051/0004-6361/202245530

Raj, M. A., Iodice, E., Napolitano, N. R., et al. 2020, A&A, 640, A137, doi: 10.1051/0004-6361/202038043

Rhee, J., Smith, R., Choi, H., et al. 2020, ApJS, 247, 45, doi: 10.3847/1538-4365/ab7377

Rudick, C. S., Mihos, J. C., Frey, L. H., & McBride, C. K. 2009, ApJ, 699, 1518, doi: 10.1088/0004-637X/699/2/1518

Rudick, C. S., Mihos, J. C., & McBride, C. K. 2011, ApJ, 732, 48, doi: 10.1088/0004-637X/732/1/48

Sommer-Larsen, J. 2006, MNRAS, 369, 958, doi: 10.1111/j.1365-2966.2006.10352.x

Spavone, M., Iodice, E., Capaccioli, M., et al. 2018, ApJ, 864, 149, doi: 10.3847/1538-4357/aad6e9

Springel, V. 2005, MNRAS, 364, 1105, doi: 10.1111/j.1365-2966.2005.09655.x

Springel, V., Pakmor, R., Zier, O., & Reinecke, M. 2021, MNRAS, 506, 2871, doi: 10.1093/mnras/stab1855

Werner, S. V., Hatch, N. A., Matharu, J., et al. 2023, MNRAS, 523, 91, doi: 10.1093/mnras/stad1410

Zhang, Y., Yanny, B., Palmese, A., et al. 2019, ApJ, 874, 165, doi: 10.3847/1538-4357/ab0dfd

Zibetti, S., White, S. D. M., Schneider, D. P., & Brinkmann, J. 2005, MNRAS, 358, 949, doi: 10.1111/j.1365-2966.2005.08817.x

Zwicky, F. 1937, ApJ, 86, 217, doi: 10.1086/143864